\documentclass
[preprint,showkeys,fleqn,prd,a4paper,11pt,tightenlines,superscriptaddress]{revtex4}%
\usepackage{amsfonts}
\usepackage{amssymb}
\usepackage{graphicx}
\usepackage{amsmath}%
\providecommand{\U}[1]{\protect\rule{.1in}{.1in}}
\begin{document}
\preprint{KEK Preprint 2015-49 CHIBA-EP-214}
\title{Confinement/deconfinement phase transition in SU(3) Yang-Mills theory in view
of dual superconductivity}
\author{Akihiro Shibata}
\affiliation{Computing Research Center, High Energy Accelerator Research Organization
(KEK), Tsukuba 305-0801, Japan}
\author{Kei-Ichi Kondo}
\affiliation{Department of Physics, Graduate School of Science, Chiba University, Chiba
263-8522, Japan}
\author{Seikou Kato}
\affiliation{Oyama National College of Technology, Oyama, Tochigi 323-0806, Japan}
\author{Toru Shinohara}
\affiliation{Department of Physics, Graduate School of Science, Chiba University, Chiba
263-8522, Japan}
\begin{abstract}
In the preceeding works, we have given a non-Abelian dual superconductivity
picture for quark confinement, and demonstrated the numerical evidences on the
lattice. In this talk, we discuss the confinement and deconfinement phase
transition at finite temperature in view of the dual superconductivity. We
investigate chromomagnetic monopole currents induced by chromoelectric flux in
both confinement and deconfinement phase by the numerical simulations on a
lattice at finite temperature, and discuss the role of the chromomagnetic
monopole in the confinement/deconfinement phase transition.

\end{abstract}
\maketitle

\section{Introduction}

The dual superconductivity is a promising mechanism for quark
confinement.\cite{DualMeisser75} In order to establish this picture, we have
to show evidences of the dual version of the superconductivity. For this
purpose, we have presented a new formulation of the Yang-Mills theory and
shown the numerical evidences on a lattice: the non-Abelian magnetic monopole
dominantly reproduces the string tension in the linear potential in
SU(3)Yang-Mills theory, and the SU(3) Yang-Mills vacuum is the type I dual
superconductor profiled by the chromoelectric flux tube and the magnetic
monopole current induced around it, which is a novel feature obtained by our
simulations.\cite{PhysRep}\cite{abeliandomSU(3)}\cite{DMeisner-TypeI2013}

In this talk, we further study the confinement and deconfinement phase
transition at finite temperature in view of the dual superconductivity. We
introduce a new formulation of the Yang-Mills theory on a lattice, and
investigate confinement/deconfinement phase transition at finite temperature
by using the new variable which extracts the dominant mode of the quark as
well as original Yang-Mills fields. We first measure the space-averaged
Polyakov-loop for each configuration and the Polyakov-loop average to
investigate the role of the new variable at finite temperature. We then
measure chromo fluxes and induced magnetic-monopole currents induced by a pair
of quark and antiquark source to investigate the dual Meissner effect. We will
demonstrate confinement/deconfinement phase transition in view of the
non-Abelian dual superconductivity picture.

\section{Gauge-link decomposition}

We introduce a new formulation of the lattice Yang-Mills theory in the minimal
option, which extracts the dominant mode of the quark confinement for $SU(3)$
Yang-Mills theory\cite{PhysRep}\cite{abeliandomSU(3)}, since we consider the
quark confinement in the fundamental representation. Let $U_{x,\mu}=X_{x,\mu
}V_{x,\mu}$ be a decomposition of the Yang-Mills link variable $U_{x,\mu}$,
where $V_{x,\mu}$ could be the dominant mode for quark confinement, and
$X_{x,\mu}$ the remainder part. The Yang-Mills field and the decomposed new
variables are transformed by full $SU(3)$ gauge transformation $\Omega_{x}$
such that $V_{x,\mu}$ is transformed as the gauge link variable and $X_{x,\mu
}$ as the site variable:
\begin{subequations}
\label{eq:gaugeTransf}%
\begin{align}
U_{x,\mu} &  \longrightarrow U_{x,\nu}^{\prime}=\Omega_{x}U_{x,\mu}%
\Omega_{x+\mu}^{\dag},\\
V_{x,\mu} &  \longrightarrow V_{x,\nu}^{\prime}=\Omega_{x}V_{x,\mu}%
\Omega_{x+\mu}^{\dag},\text{ \ }X_{x,\mu}\longrightarrow X_{x,\nu}^{\prime
}=\Omega_{x}X_{x,\mu}\Omega_{x}^{\dag}.
\end{align}
The decomposition is given by solving the defining equation:
\end{subequations}
\begin{subequations}
\label{eq:DefEq}%
\begin{align}
&  D_{\mu}^{\epsilon}[V]\mathbf{h}_{x}:=\frac{1}{\epsilon}\left[  V_{x,\mu
}\mathbf{h}_{x+\mu}-\mathbf{h}_{x}V_{x,\mu}\right]  =0,\label{eq:def1}\\
&  g_{x}:=e^{i2\pi q/3}\exp(-ia_{x}^{0}\mathbf{h}_{x}-i\sum\nolimits_{j=1}%
^{3}a_{x}^{(j)}\mathbf{u}_{x}^{(j)})=1,\label{eq:def2}%
\end{align}
where $\mathbf{h}_{x}$ is an introduced color field $\mathbf{h}_{x}=\xi
_{x}(\lambda^{8}/2)\xi_{x}^{\dag}$ $\in\lbrack SU(3)/U(2)]$ with $\lambda^{8}$
being the Gell-Mann matrix and $\xi_{x}$ an $SU(3)$ group element. The
variable $g_{x}$ is an undetermined parameter from Eq.(\ref{eq:def1}),
$\mathbf{u}_{x}^{(j)}$ 's are $su(2)$-Lie algebra valued, and $q_{x}$ has an
integer value $\ 0,1,2$. These defining equations can be solved exactly
\cite{exactdecomp}, and the solution is given by
\end{subequations}
\begin{subequations}
\label{eq:decomp}%
\begin{align}
X_{x,\mu} &  =\widehat{L}_{x,\mu}^{\dag}\det(\widehat{L}_{x,\mu})^{1/3}%
g_{x}^{-1},\text{ \ \ \ }V_{x,\mu}=X_{x,\mu}^{\dag}U_{x,\mu}=g_{x}\widehat
{L}_{x,\mu}U_{x,\mu},\\
\widehat{L}_{x,\mu} &  =\left(  L_{x,\mu}L_{x,\mu}^{\dag}\right)
^{-1/2}L_{x,\mu},\text{ }\\
L_{x,\mu} &  =\frac{5}{3}\mathbf{1}+\frac{2}{\sqrt{3}}(\mathbf{h}_{x}%
+U_{x,\mu}\mathbf{h}_{x+\mu}U_{x,\mu}^{\dag})+8\mathbf{h}_{x}U_{x,\mu
}\mathbf{h}_{x+\mu}U_{x,\mu}^{\dag}\text{ .}%
\end{align}
Note that the above defining equations correspond to the continuum version

The decomposition is uniquely obtained as the solution (\ref{eq:decomp}) of
Eqs.(\ref{eq:DefEq}), if color fields$\{\mathbf{h}_{x}\}$ are obtained. To
determine the configuration of color fields, we use the reduction condition to
formulate the new theory written by new variables ($X_{x,\mu}$,$V_{x,\mu}$)
which is equipollent to the original Yang-Mills theory. Here, we use the
reduction functional:
\end{subequations}
\begin{equation}
F_{\text{red}}[\mathbf{h}_{x}]=\sum_{x,\mu}\mathrm{tr}\left\{  (D_{\mu
}^{\epsilon}[U_{x,\mu}]\mathbf{h}_{x})^{\dag}(D_{\mu}^{\epsilon}[U_{x,\mu
}]\mathbf{h}_{x})\right\}  , \label{eq:reduction}%
\end{equation}
and then color fields $\left\{  \mathbf{h}_{x}\right\}  $ are obtained by
minimizing the functional (\ref{eq:reduction}).

\section{Lattice result}

We generate the Yang-Mills gauge field configurations (link variables)
$\{U_{x,\mu}\}$ for the standard Wilson action. We prepare data sets for
finite temperature on the lattice of size $L^{3}\times N_{T}$ at finite
temperature by using the pseudo heat bath algorithm. For the fixed spatial
size $L$ and the temporal size $N_{T}$: $L=24$, $N_{T}=6$: the temperature
varies by changing the inverse gauge coupling constant $\beta=2N_{c}/g^{2}$
($N_{c}=3)$: $\beta=5.8$, $5.9$, $6.0$, $6.1$, $6.2$, $6.3$ . In our
simulations, we use the cold start to obtain the real-valued Polyakov loop
average $\langle P\rangle$ at high temperature, see Fig.~\ref{Fig:PLP}. We
thermalized 8000 sweeps, and we used 500 configurations for measurements.

We perform the decomposition of the gauge link variable $U_{x,\mu}=X_{x,\mu
}V_{x,\mu}$ by using the formula (\ref{eq:decomp}) given in the previous
section, after the color-field configuration $\{\mathbf{h}_{x}\}$ is obtained
by solving the reduction condition of minimizing the functional
(\ref{eq:reduction}) for each set of the gauge field configurations
$\{U_{x,\mu}\}$. In the measurement of the Polyakov loop average and the
Wilson loop average defined below, we apply the APE smearing technique to
reduce noises.

\subsection{Polyakov-loop average in the confinement/deconfinement transition}

\begin{figure}[ptb]
\begin{center}
\includegraphics[
height=5cm, angle=270]
{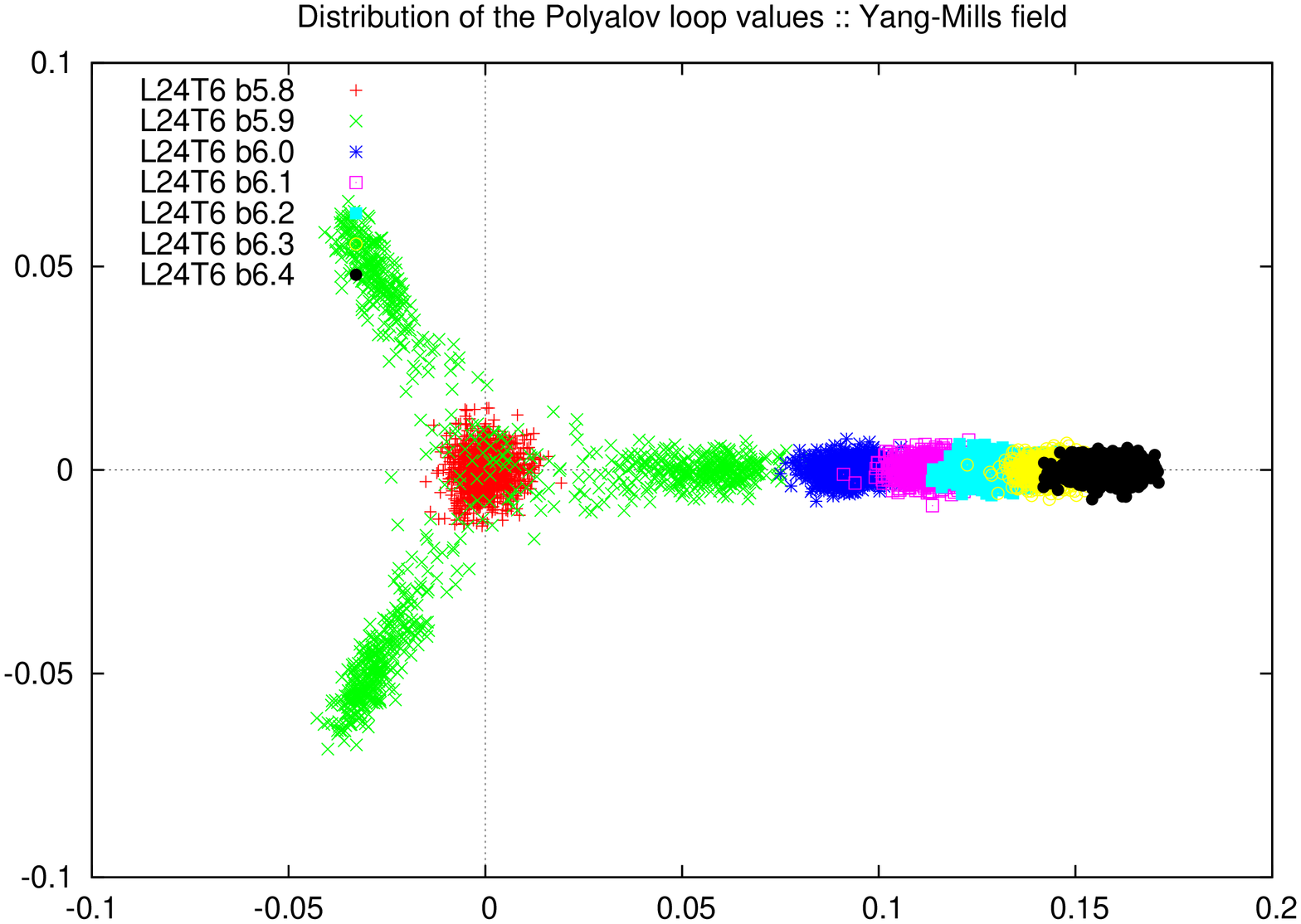} \includegraphics[
height=5cm, angle=270]
{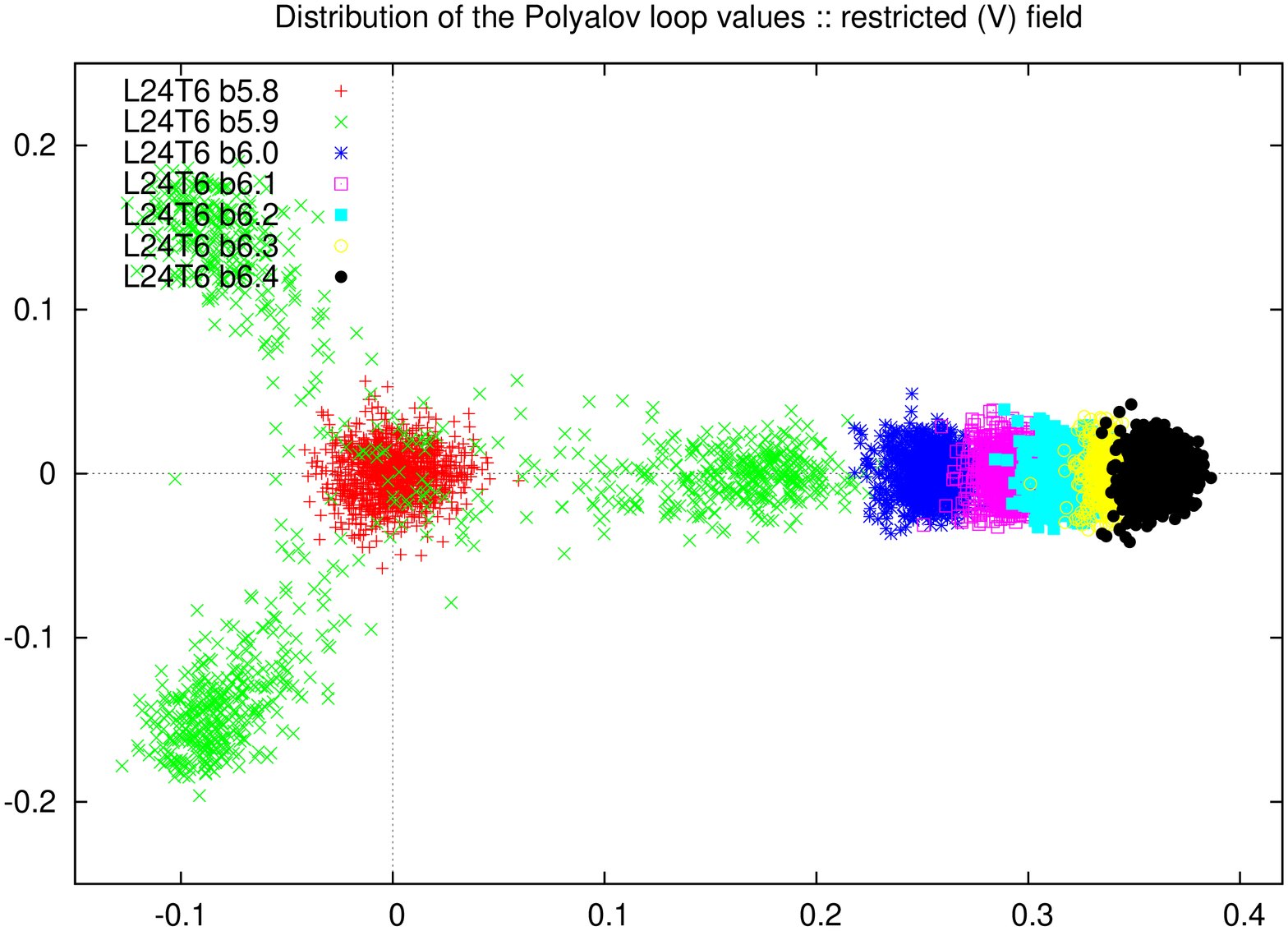} \includegraphics[
height=5cm, angle=270]
{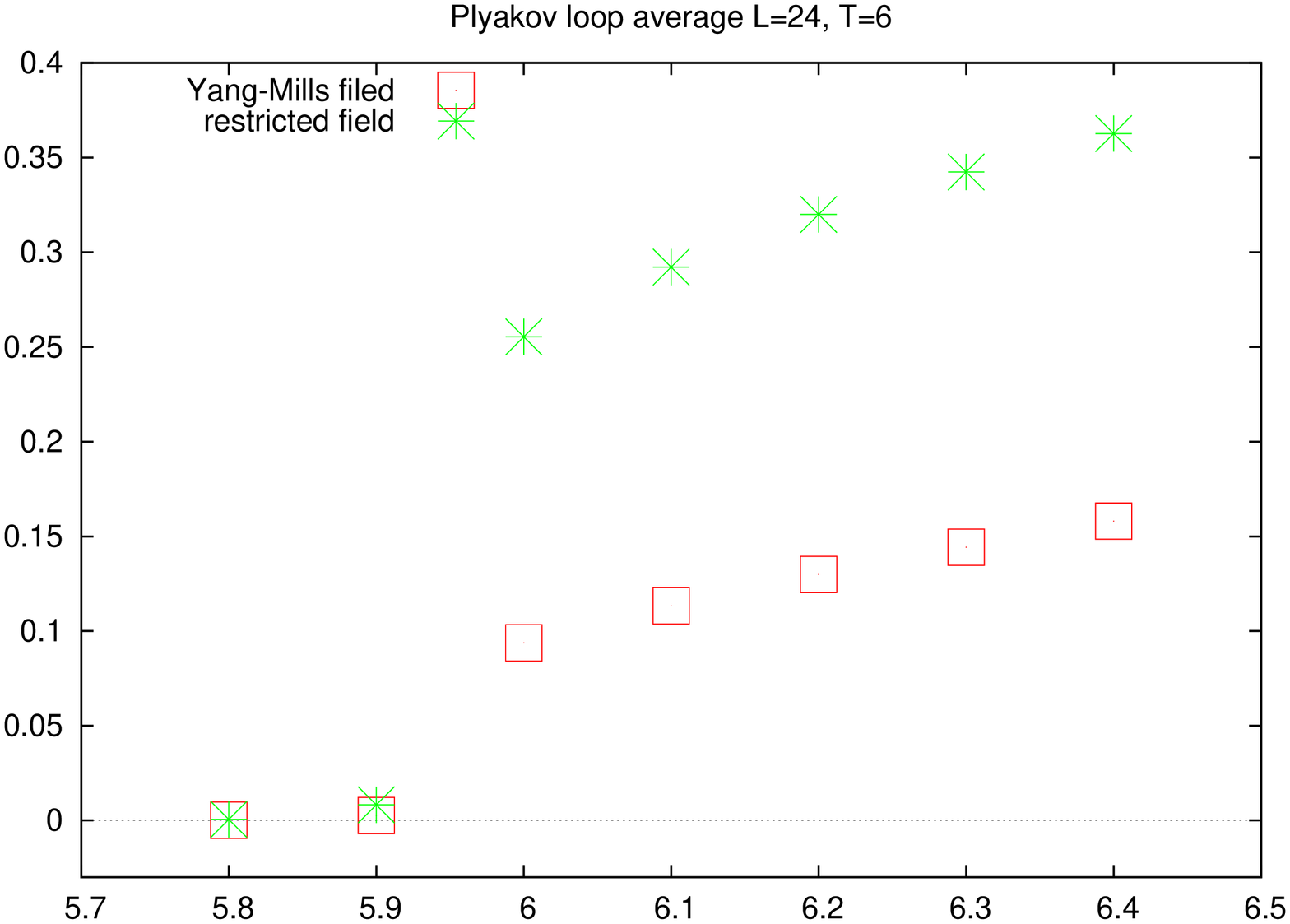}
\end{center}
\caption{ {}The distribution of the space-averaged Polyakov loop on the
complex plane: (Left) $P_{U}$ obtained from a set of the original gauge field
configurations, (Center) $P_{V}$ from the restricted field. (Right) The
Polyakov-loop average versus temperature. The red and green represent the
Polyakov-loop average for the original field $U$ and the restricted field $V$,
respectively. }%
\label{Fig:PLP}%
\end{figure}First, we measure the Polyakov-loop average which is a
conventional order parameter for detecting the confinement and deconfinement
phase transition in the pure Yang-Mills theory. We define the space-averaged
Polyakov loop (i.e., the value of the Polyakov loop which is averaged over the
space volume) for a set of the original gauge field configurations
$\{U_{x,\mu}\}$ and the restricted gauge field ($V$-field) configurations
$\{V_{x,\mu}\}$:
\begin{equation}
P_{U}:=L^{-3}\sum_{\vec{x}}\mathrm{tr}\left(  \prod\nolimits_{t=1}^{N_{T}%
}U_{(\vec{x},t),4}\right)  ,\quad P_{V}:=L^{-3}\sum_{\vec{x}}\mathrm{tr}%
\left(  \prod\nolimits_{t=1}^{N_{T}}V_{(\vec{x},t),4}\right)  .\label{eq:PLP}%
\end{equation}
Left and center panel of Fig.~\ref{Fig:PLP} show the plots of $P_{U}$ (left
panel) and $P_{V}$ (center panel) on the complex plane measured from the
original gauge field configurations and the restricted gauge field
configurations respectively. Notice that the Polyakov loop average is in
general complex-valued for the $SU(3)$ group.

Then, we measure the Polyakov-loop average $\left\langle P_{U}\right\rangle $
and $\left\langle P_{V}\right\rangle $ obtained by averaging the
space-averaged Polyakov loop over the total sets of the original gauge field
configurations and the restricted gauge field configurations respectively.
Note that the symbol $\left\langle \mathcal{O}\right\rangle $ denotes the
average of the operator $\mathcal{O}$ over the space and the ensemble of the
configurations. Right panel of Fig.~\ref{Fig:PLP} shows the result for the
Polyakov loop average for various temperature ($\beta$). We find that the
behavior of $\left\langle P_{U}\right\rangle $ and $\left\langle
P_{V}\right\rangle $ give the same critical temperature for the phase
transition separating the low-temperature confined phase characterized by the
vanishing Polyakov loop average $\left\langle P_{U}\right\rangle =\left\langle
P_{V}\right\rangle =0$ from the high-temperature deconfined phase
characterized by the non-vanishing Polyakov loop average $\left\langle
P_{U}\right\rangle \neq0$ and $\left\langle P_{V}\right\rangle \neq0$.

\subsection{Chromo-flux tube at finite temperature}

\begin{figure}[tb]
\begin{center}
\includegraphics[height=4cm]
{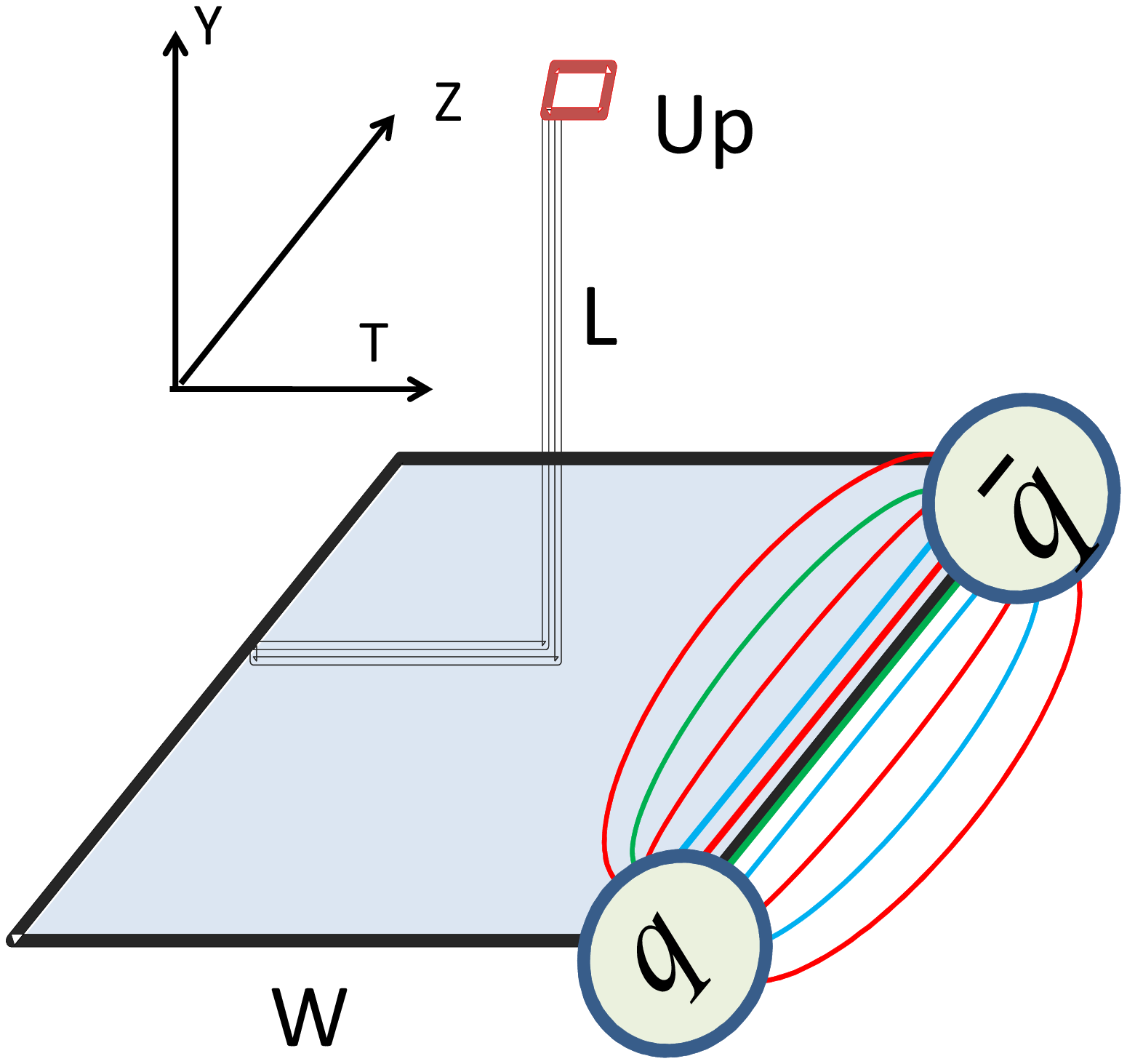} \ \ \ \includegraphics[height=4cm]
{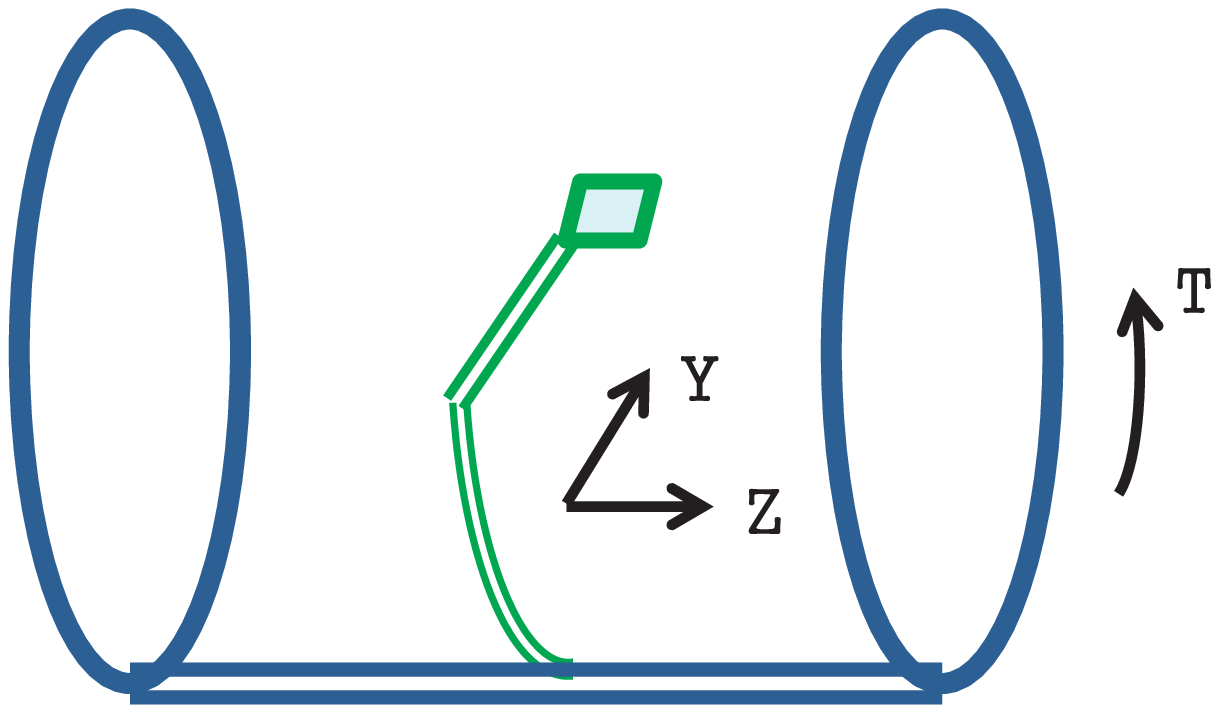} \ \ \
\end{center}
\caption{The setup of the measurement of the chromo-field strength
distribution. (Left) The gauge-invariant operator $\mathrm{tr}(WLU_{p}L^{\dag
})$ between a plaquette $U_{p}$ and the Wilson loop $W$. (Right) Measurement
of the chromo-flux at finite temperature. }%
\label{fig:measure2}%
\end{figure}

We proceed to investigate the non-Abelian dual Meissner effect at finite
temperature. For this purpose, we measure the chromo-flux at finite
temperature created by a quark-antiquark pair which is represented by the
maximally extended Wilson loop $W$ defined in Fig.\ref{fig:measure2}. The
chromo-field strength, i.e., the field strength of the chromo flux at the
position $P$ is measured by using a plaquette variable $U_{p}$ as the probe
operator for the field strength. See Fig.~\ref{fig:measure2}. We use the same
gauge-invariant correlation function as that used at zero temperature
\cite{Giacomo}:
\begin{equation}
\rho_{_{U_{P}}}:=\frac{\left\langle \mathrm{tr}\left(  WLU_{p}L^{\dag}\right)
\right\rangle }{\left\langle \mathrm{tr}\left(  W\right)  \right\rangle
}-\frac{1}{N_{c}}\frac{\left\langle \mathrm{tr}\left(  U_{p}\right)
\mathrm{tr}\left(  W\right)  \right\rangle }{\left\langle \mathrm{tr}\left(
W\right)  \right\rangle },\label{eq:Op}%
\end{equation}
where $L$ is the Wilson line connecting the source $W$ and the probe $U_{p}$
needed to obtain the gauge-invariant result. Note that $\rho_{_{U_{P}}}$ is
sensitive to the field strength rather than the disconnected one. Notice that
the setup in the right figure of Fig.~\ref{fig:measure2} is different from the
correlation function for a pair of the Polyakov loop $P(z=0)$ and the
anti-Polyakov loop $\bar{P}(z=R)$ where each Polyakov loop is defined by the
corresponding closed loop which is obtained by identifying the end points at
$\tau=0$ and $\tau=1/T$ by the periodic boundary condition, and the probe
$U_{P}$ is attached to one the Polyakov loop or both the Polyakov and
anti-Polyakov loops. Such operator was recently used to measure the
chromo-flux in the work \cite{P.Cea-L.Lenardo2014}. It should be remarked that
the Polyakov loop correlation function, $\left\langle P_{U}(\vec{x}%
)P_{U}^{\ast}(\vec{y})\right\rangle \simeq e^{-F_{q\bar{q}}/T}$, proportional
to the partition function in the presence of a quark at $\vec{x}$ and an
anti-quark at $\vec{y}$ is decomposed into the singlet and the adjoint
combinations in the color space. Furthermore, the decomposed component is
gauge-dependent and thus should be taken with care.\cite{YHM05}\cite{BW79} In
sharp contrast to this fact, the potential obtained from the Wilson loop is
the color singlet in the gauge-independent way.

\begin{figure}[ptb]
\begin{center}
\includegraphics[
height=5cm, angle=270]
{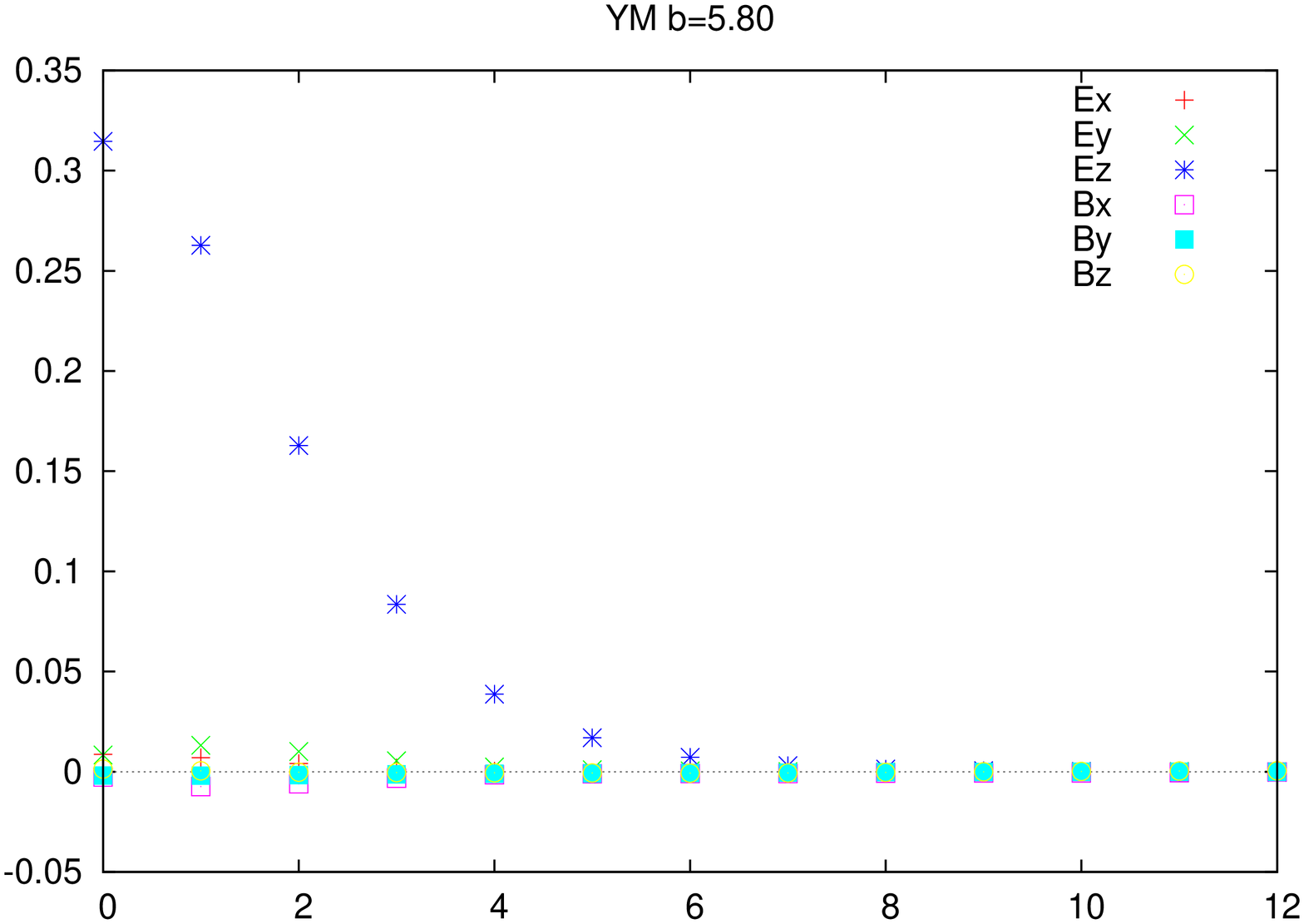} \includegraphics[
height=5cm, angle=270]
{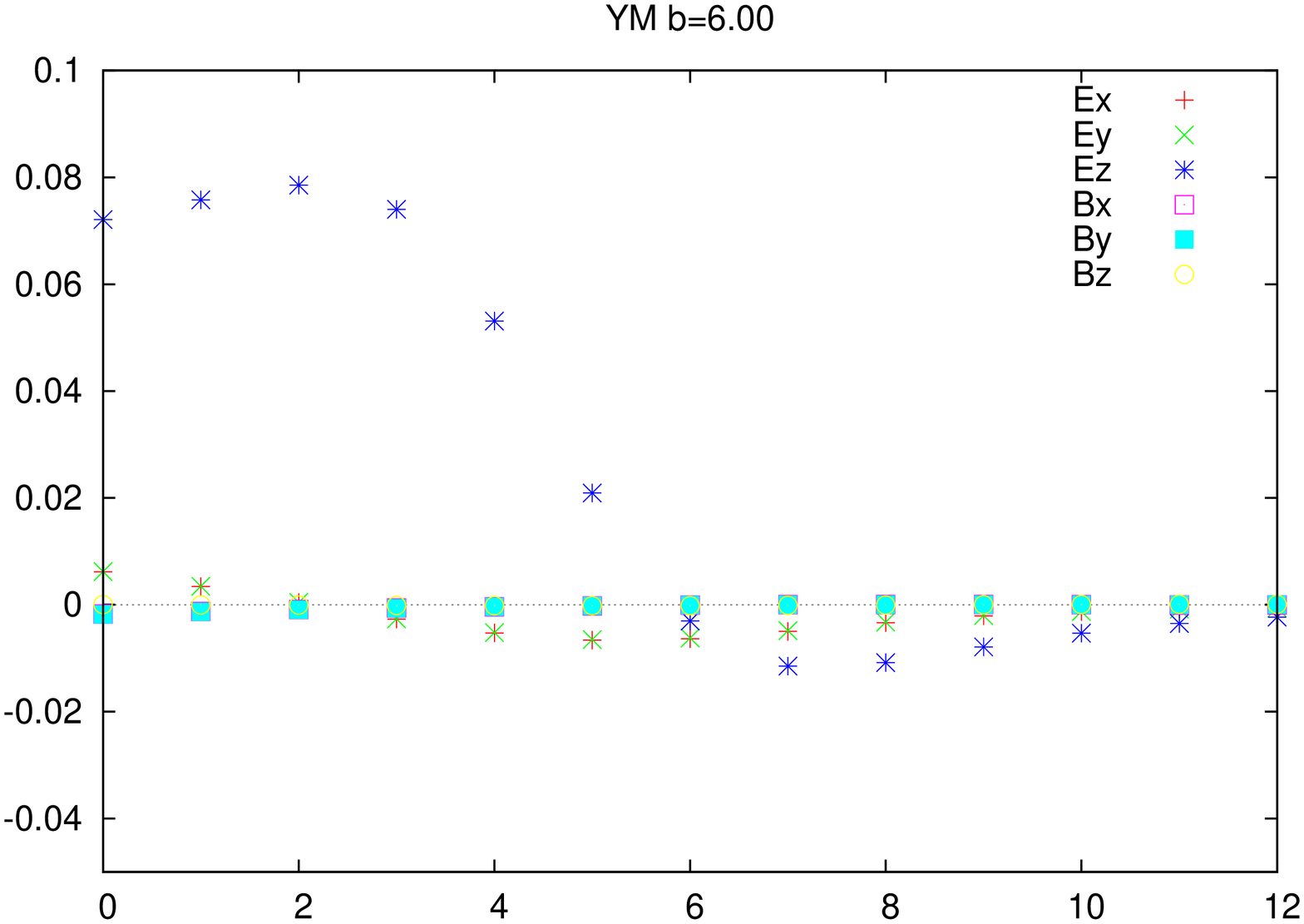} \includegraphics[
height=5cm, angle=270]
{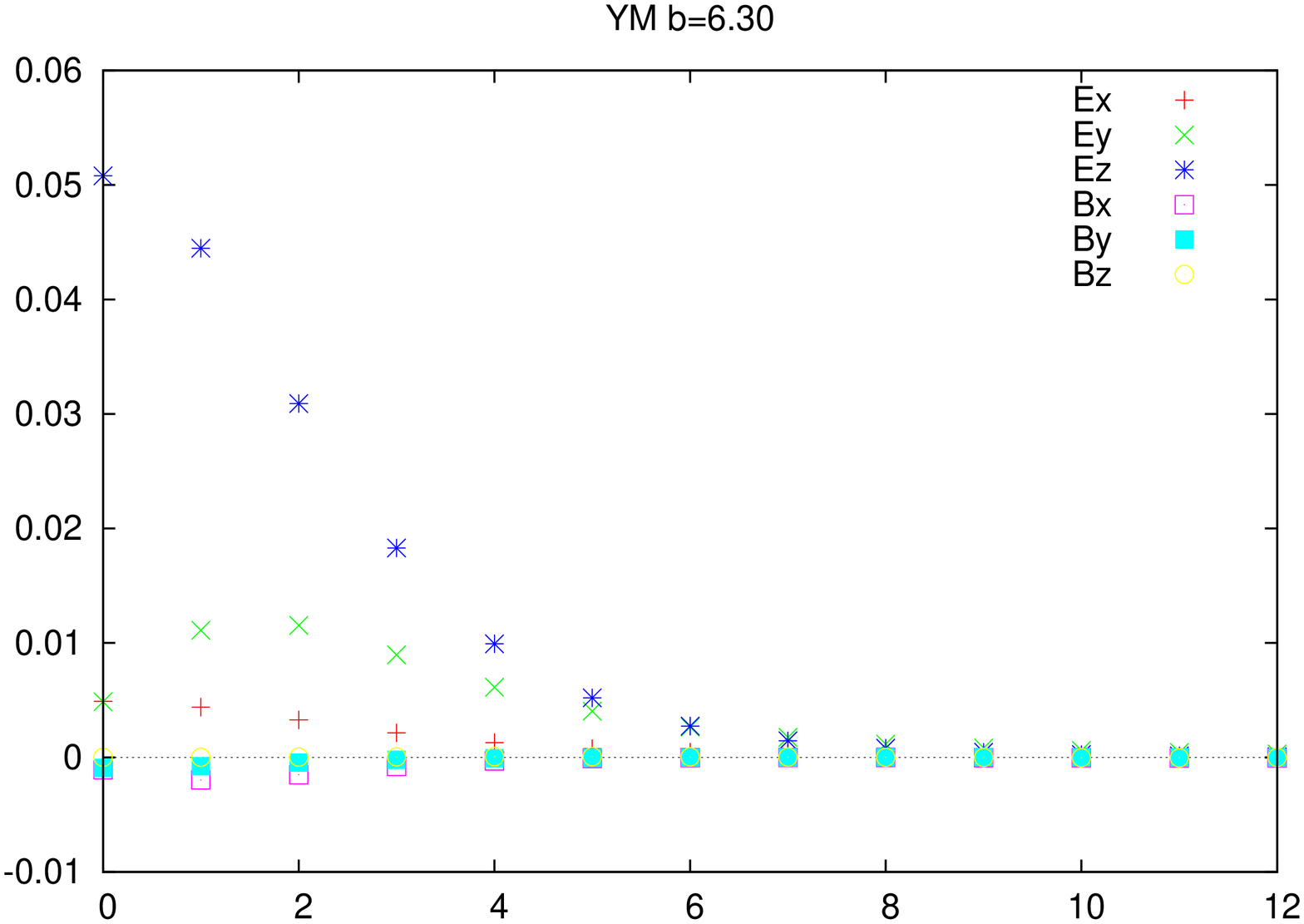}
\end{center}
\caption{ Chromo-field strength of the original Yang-Mills field created by
the quark-antiquark pair at finite temperature: (left) $\beta=5.80$ (center)
$\beta=6.00$ (right) $\beta=6.30$ on the lattice $24^{3}\times6$. }%
\label{fig:flux-YM}%
\end{figure}

\begin{figure}[ptb]
\begin{center}
\includegraphics[
height=5cm, angle=270]
{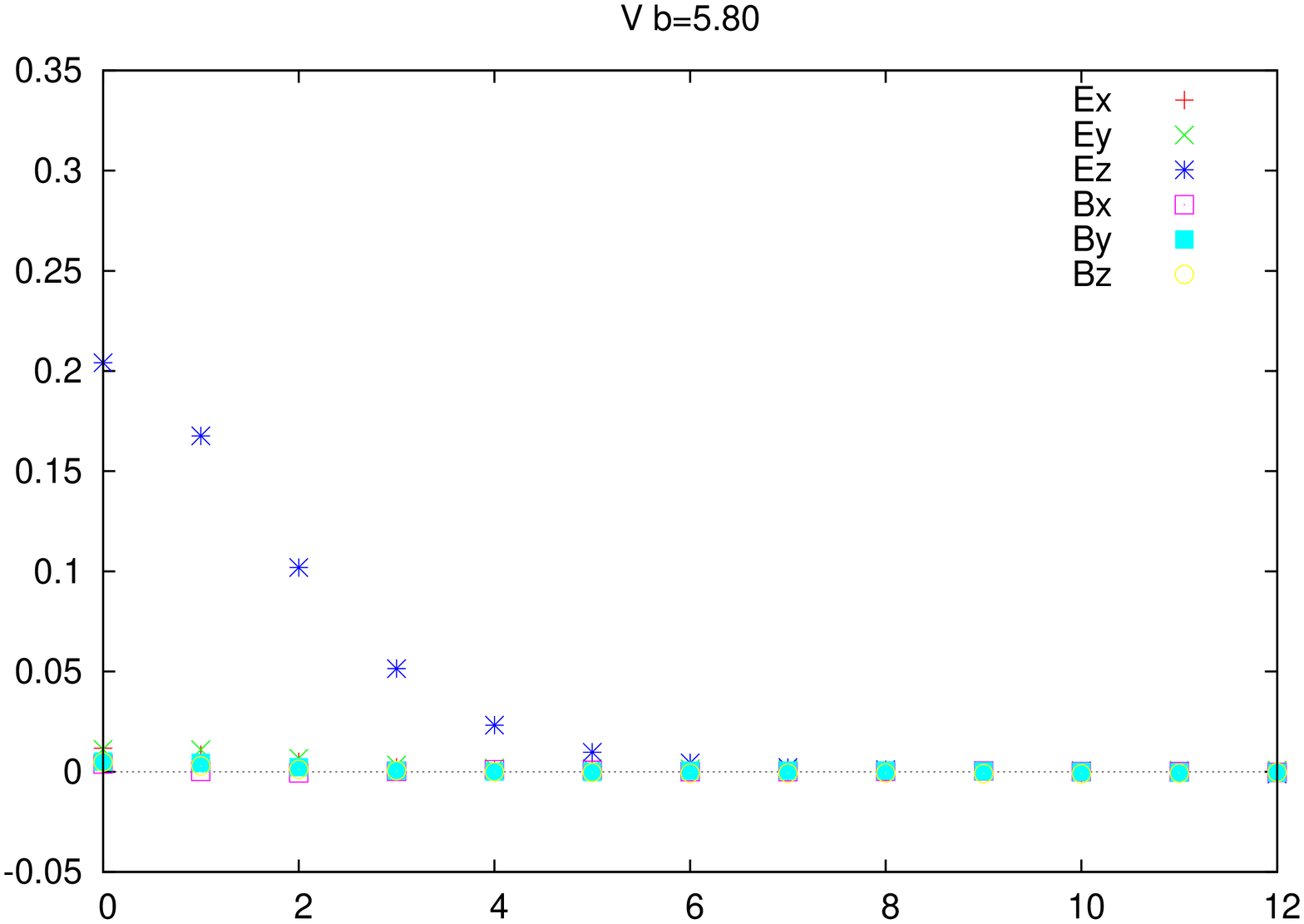}\includegraphics[
height=5cm, angle=270]
{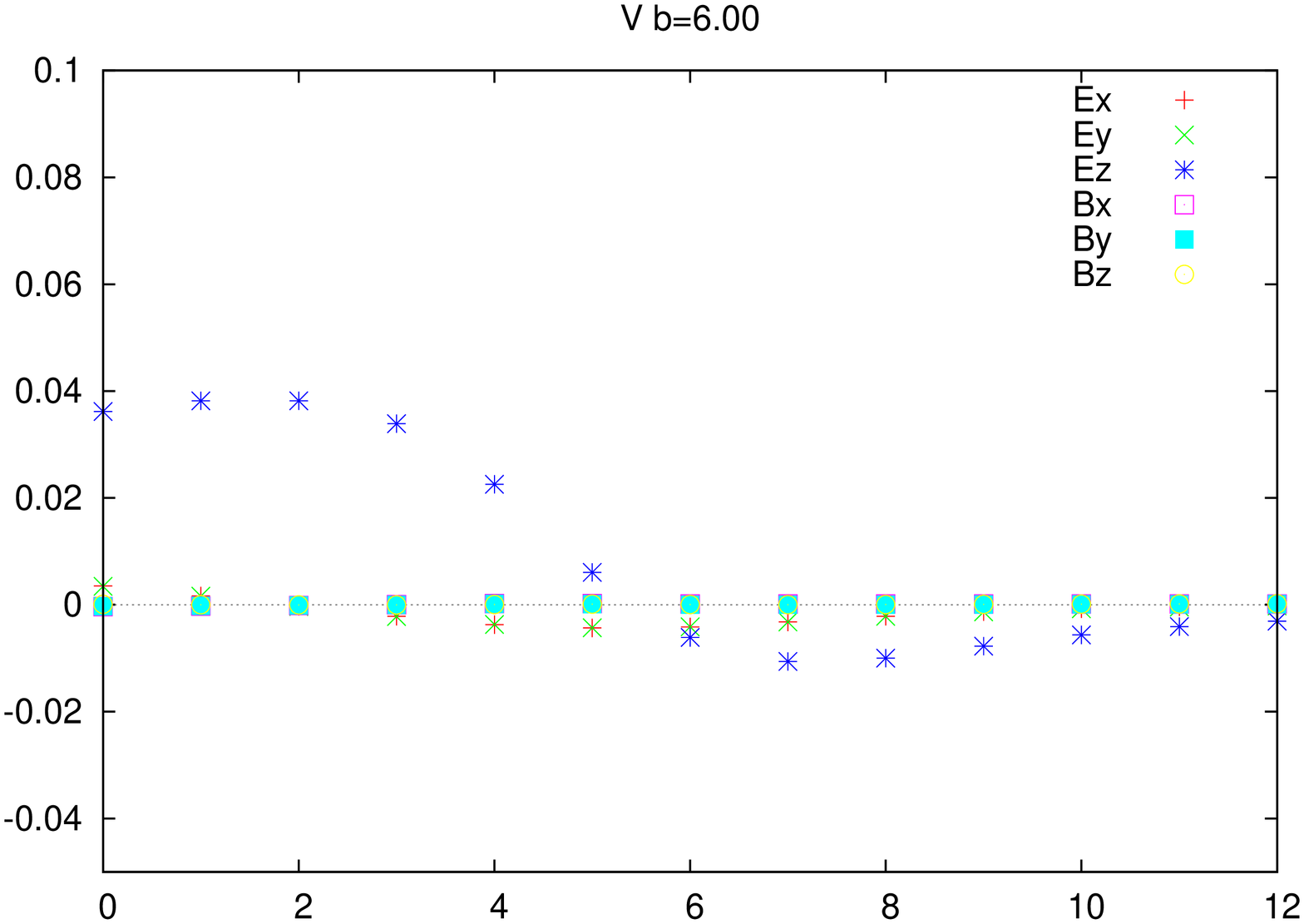}\includegraphics[
height=5cm, angle=270]
{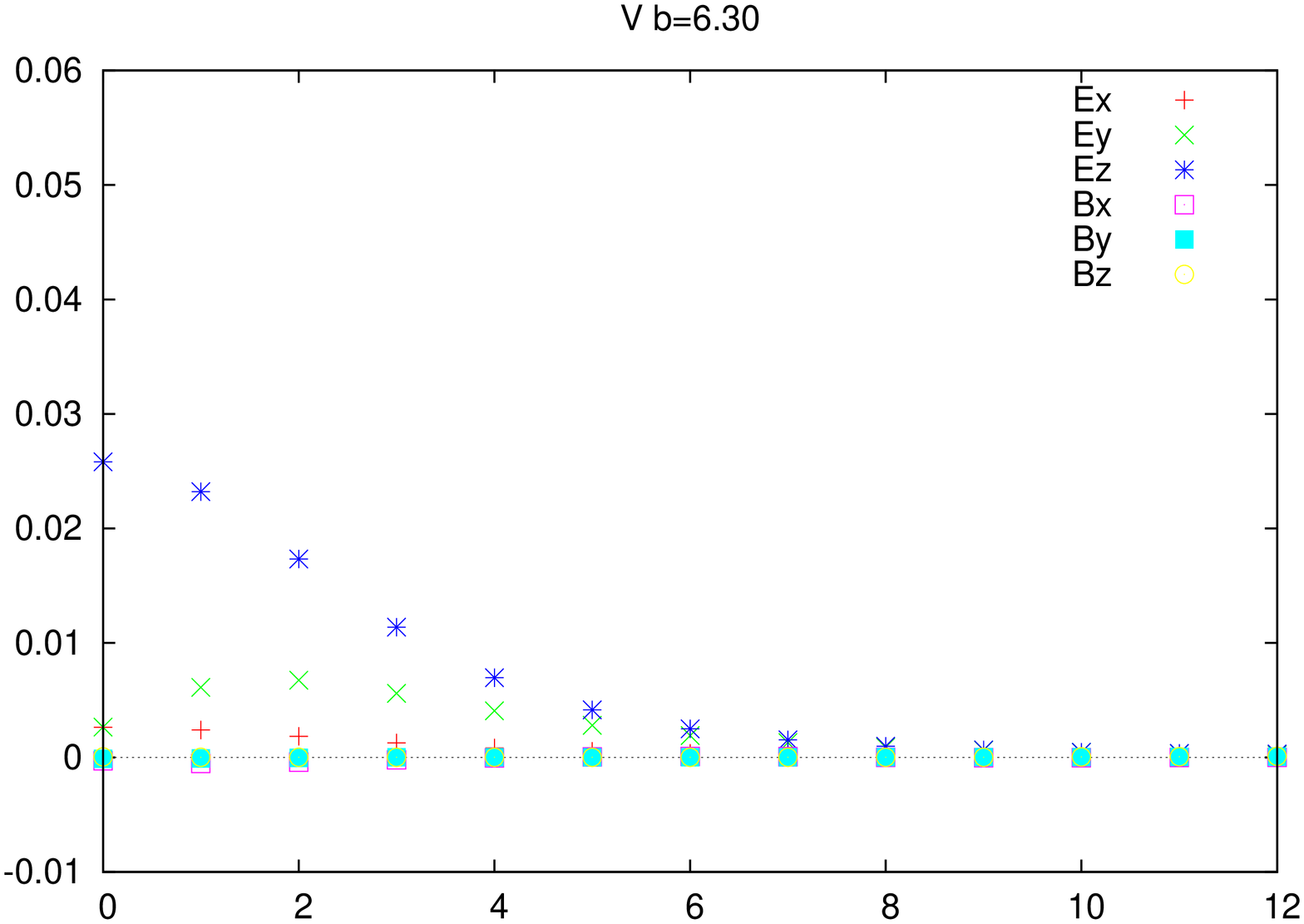}
\end{center}
\caption{ Chromo-field strength of the restricted field created by the
quark-antiquark pair at finite temperature: (left) $\beta=5.80$ (center)
$\beta=6.00$ (right) $\beta=6.30$ on the lattice $24^{3}\times6$. }%
\label{fig:flux-V0}%
\end{figure}Figure \ref{fig:flux-YM} and \ref{fig:flux-V0} show the results of
the measurement of chromo-field strength at different temperatures obtained
from the data set for the original Yang-Mills field and the restricted field
($V$-field), respectively. In the low temperature confined phase $T<T_{c}$
(see left panels of Fig.\ref{fig:flux-YM} and Fig.\ref{fig:flux-V0}), we
observe that only the $E_{z}$ component of the chromoelectric flux tube, i.e.
the flux in the direction connecting a quark and antiquark pair is
non-vanishing, and that the other components take vanishing values. This is
consistent with the result obtained by Cea et al. \cite{P.Cea-L.Lenardo2014},
although they use the different operator for measuring the flux. In the high
temperature deconfined phase $T>T_{c}$ (see right panels of
Fig.~\ref{fig:flux-YM} and Fig.~\ref{fig:flux-V0}), we observe the
non-vanishing $E_{y}$ component in the chromoelectric flux, which means no
more squeezing of the chromoelectric flux tube. This shows the disappearance
of the dual Meissner effect in the high temperature deconfined phase.

\subsection{Magnetic--monopole current and dual Meissner effect at finite
temperature}

\begin{figure}[ptb]
\begin{center}
\includegraphics[height=5cm]{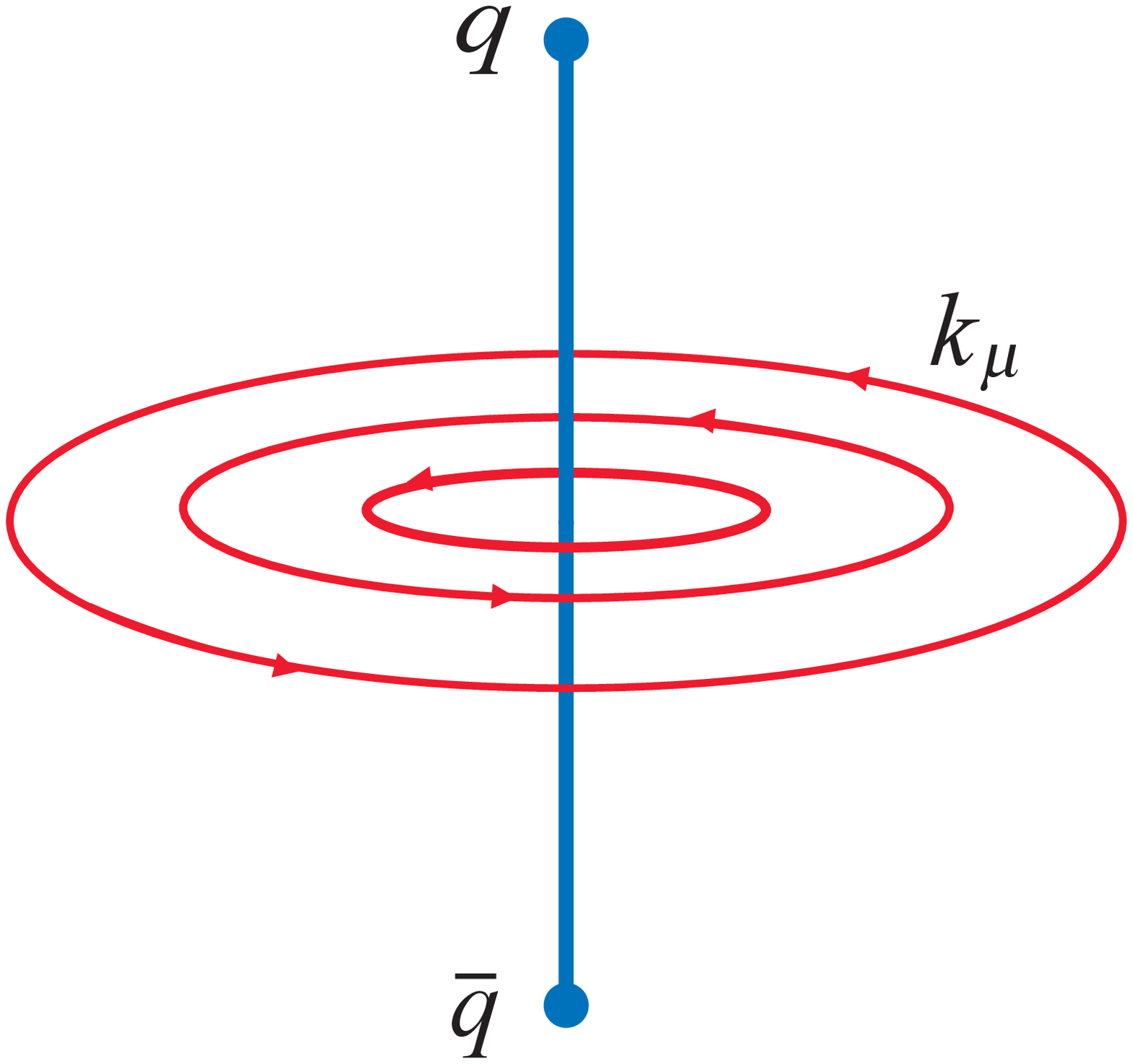} \hspace{15mm}
\includegraphics[height=6cm, angle=270, origin=br]
{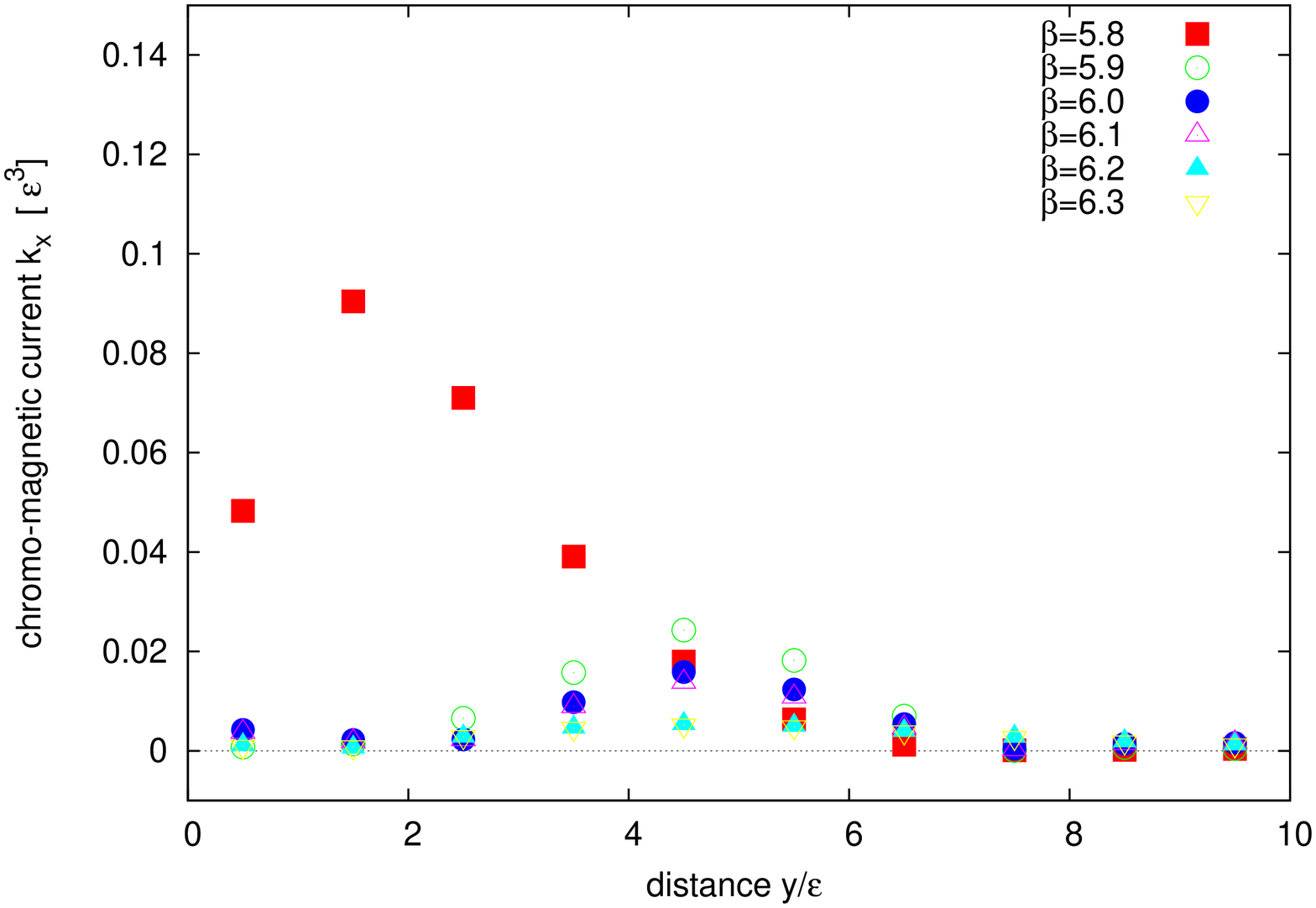}
\end{center}
\caption{ (Left) Sketch of the induced magnetic-monopole current by a pair of
quark and anti-quark. (Right) The magnitude $\sqrt{k_{\mu}k_{\mu}}$ of the
induced magnetic current $k_{\mu}$ around the flux tube connecting the
quark-antiquark pair as a function of the distance $y$ from the axis $z$ for
various values of $\beta$, i.e., temperature. }%
\label{fig:m_current}%
\end{figure}

Finally, we investigate the dual Meissner effect by measuring the
magnetic--monopole current $k$ induced around the chromo-flux tube created by
the quark-antiquark pair. We use the the magnetic-monopole current $k$ defined
by
\begin{equation}
k_{\mu}(x)=\frac{1}{2}\epsilon_{\mu\nu\alpha\beta}\left(  F[V]_{\alpha\beta
}(x+\hat{\nu})-F[V]_{\alpha\beta}(x)\right)  . \label{eq:m_current}%
\end{equation}
Note that the magnetic--monopole current (\ref{eq:m_current}) must vanish due
to the Bianchi identity, if there exist no singularity in the gauge potential.
We show that the magnetic--monopole current defined in this way can be the
order parameter for the confinement/deconfinement phase transition, as
suggested from the dual superconductivity hypothesis. Figure
\ref{fig:m_current} shows the result of the measurements of the magnitude
$\sqrt{k_{\mu}k_{\mu}}$
%the $x$ component $k_x$
of the induced magnetic current $k_{\mu}$ obtained according to
(\ref{eq:m_current}). We observe the appearance and disappearance of the
magnetic monopole current in the low temperature phase and high temperature
phase, respectively.

\section{Summary and outlook}

Using a new formulation of the Yang-Mills theory on a lattice, we have
examined the confinement/deconfinement phase transition and the (non-Abelian)
dual superconductivity in the $SU(3)$ Yang-Mills theory at finite temperature.
The reformulation enables one to extract the dominant mode for quark
confinement. Indeed, we have extracted the restricted field ($V$-field) from
the original Yang-Mills field which plays a dominant role in confining quark
in the fundamental representation at finite temperature.

First, we have given the numerical evidences for the restricted field
dominance in the Polyakov loop average $P$ in the sense that the Polyakov loop
average $P_{V}$ written in terms of the restricted field $V$ gives the same
critical temperature $T_{c}$ as that detected by the Polyakov loop average
$P_{U}$ written in terms of the original gauge field $U$: $P=0$ for $T<T_{c}$
and $P \ne0$ for $T>T_{c}$.

However, the Polyakov loop average cannot be the direct signal of the dual
Meissner effect or magnetic monopole condensation. Therefore, it is important
to find an order parameter which enables one to detect the dual Meissner
effect directly.
%and to investigate whether or not such an order parameter gives the same critical temperature as that detected by the Polyakov loop average in view of the dual Meissner effect.

In view of these, we have measured the chromoelectric and chromomagnetic flux
for both the original field and the restricted field. In the low--temperature
confined phase $T<T_{c}$, we have obtained the numerical evidences of the dual
Meissner effect in the $SU(3)$ Yang-Mills theory, i.e., the squeezing of the
chromoelectric flux tube created by a quark-antiquark pair and the associated
magnetic--monopole current induced around the flux tube. In the
high--temperature deconfined phase $T>T_{c}$, on the other hand, we have
observed the disappearance of the dual Meissner effect, no more squeezing of
the chromoelectric flux tube detected by non-vanishing $E_{y}$ component in
the chromoelectric flux and the vanishing of the magnetic-monopole current
associated with the chromo-flux tube. These results are also obtained by the
restricted field alone. Therefore, we have confirmed the restricted field
dominance in the dual Meissner effect even at finite temperature. Thus, we
have given the evidences that the confinement/deconfinement phase transition
is caused by appearance/disappearance of the non-Abelian dual superconductivity.

\subsection*{Acknowledgement}

This work is supported by Grant-in-Aid for Scientific Research (C) 24540252
and 15K05042 from Japan Society for the Promotion Science (JSPS), and in part
by JSPS Grant-in-Aid for Scientific Research (S) 22224003. The numerical
calculations are supported by the Large Scale Simulation Program No.13/14-23
(2013-2014) and No.14/15-24 (2014-2015) of High Energy Accelerator Research
Organization (KEK).

\end{document}